\documentclass[aps,prl,twocolumn,showpacs,superscriptaddress,showkeys,bm,amsmath,amssymb]{revtex4-1}
\usepackage{graphicx}
\usepackage{xcolor}
\usepackage{soul}
\usepackage{dcolumn}

\begin{document}
\title{Searching for Invisible Axion Dark Matter with an 18\,T Magnet Haloscope}
\newcommand{\KAIST}{\affiliation{Department of Physics, Korea Advanced Institute of Science and Technology, Daejeon, 34141, Korea}}
\newcommand{\SNU}{\affiliation{Department of Physics and Astronomy, Seoul National University, Seoul 08826, Korea}}
\newcommand{\IBS}{\affiliation{Center for Axion and Precision Physics Research, Institute for Basic Science, Daejeon, 34051, Korea}} 

\author{Youngjae Lee}\KAIST
\author{Byeongsu Yang}\IBS
\author{Hojin Yoon}\KAIST
\author{Moohyun Ahn}\SNU
\author{Heejun Park}\IBS
\author{Byeonghun Min}\IBS
\author{DongLak Kim}\IBS
\author{Jonghee Yoo}\KAIST\IBS\SNU

\date{\today}
\begin{abstract}
    We report the first search results for axion dark matter using an 18\,T high-temperature superconducting magnet haloscope. The scan frequency ranges from 4.7789 to 4.8094\,GHz. No significant signal consistent with the Galactic halo dark matter axion is observed. The results set the best upper bound of axion-photon-photon coupling ($g_{a\gamma\gamma}$) in the mass ranges of 19.764 to 19.771\,$\mu$eV (19.863 to 19.890\,$\mu$eV) at 1.5$\times|g_{a\gamma\gamma}^{\text{KSVZ}}|$ (1.7$\times|g_{a\gamma\gamma}^{\text{KSVZ}}|$), and 19.772 to 19.863\,$\mu$eV at 2.7 $\times|g_{a\gamma\gamma}^{\text{KSVZ}}|$ with 90\% confidence level, respectively. This remarkable sensitivity in the high mass region of dark matter axion is achieved by using the strongest magnetic field among the existing haloscope experiments and realizing a low-noise amplification of microwave signals using a Josephson parametric converter. 
\end{abstract}
\maketitle

\par Astrophysical observations indicate that the Universe is dominantly filled with unknown dark components. Probing dark matter is one of the most prominent subjects in modern particle and astroparticle physics\,\cite{PhysRevD.98.030001}. Axions have been postulated to solve the strong-{\it CP} problem in quantum chromodynamics. The Peccei-Quinn {\it U}$_{\rm PQ}(1)$ symmetry breaking mechanism was suggested as a solution to the problem\,\cite{Peccei:1977hh,Weinberg:1977ma,Wilczek:1977pj}. Two prominent axion models were considered as benchmarks: a {\it hadronic axion} model (Kim-Shifman-Vainshtein-Zakharov, KSVZ)\,\cite{Kim:1979if,Shifman:1979if} and a {\it fermionic axion} model (Dine-Fischler-Srednicki-Zhitnitsky, DFSZ) \cite{Dine:1981rt,Zhitnitsky:1980tq}. Couplings of these axions to ordinary matter evade almost all current experimental bounds, hence called {\it invisible axions}. Owing to these properties, axions are excellent candidates for dark matter which would have been produced during the big bang\,\cite{Planck, Abbott:1982af,DINE1983137,Preskill:1982cy,PhysRevLett.50.925,Abbott:1982af,DINE1983137,Sikivie:2006ni,PhysRevD.78.083507}. Especially in the case of {\it U}$_{\rm PQ}(1)$ symmetry breaking after the inflation, cosmological constraints suggest that the axion mass is on the scale of micro-electron-volt or higher\,\cite{Bonati2016,PhysRevD.92.034507,Borsanyi2016,PhysRevLett.118.071802,PhysRevD.96.095001,PETRECZKY2016498}. Therefore, probing the axion dark matter in these mass scales is highly encouraged.

\par Haloscopes are regarded as the most compelling detector technology to date for a dark matter axion search\,\cite{Sikivie:1983ip,Hagmann1998a,Asztalos:2003px,Asztalos:2009yp,Du:2018uak,Braine:2019fqb,ADMX:2021nhd,Brubaker:2016ktl,Backes2021a,Lee:2020cfj,Jeong:2020cwz,Kwon2021a}. A typical axion haloscope deploys a resonant radio-frequency (rf) cavity in the center of a strong solenoid magnet. The local dark matter axions are expected to couple to the applied magnetic field in the cavity and convert to microwave photons. These microwaves can be read out using rf antenna technologies. The signal power of these rf photons is given by
\begin{eqnarray}
P_{a} = g^2_{a\gamma\gamma}\Big(\dfrac{\rho_a}{m_a}\Big)B^2 V C \dfrac{\beta}{(1+\beta)^2} Q_0,
\end{eqnarray}
\noindent where the axion-photon-photon coupling ($g_{a\gamma\gamma}$) and the mass of the axion ($m_a$) are unknown parameters that need to be experimentally determined, $\rho_a$ is the local dark matter density (0.45\,GeV/{\it cc}), $B$ is the external magnetic field, $V$ is the effective volume of the resonant cavity, $C$ is the cavity form factor, $Q_0$ is the unloaded quality factor of the cavity, and $\beta$ is the coupling coefficient of the rf antenna to the cavity. The scan speed of the dark matter axion is given as $dm_a/dt \propto B^4 V^2Q_L/T_S^2$, where  $Q_L = Q_0 / (1+\beta)$ is the loaded quality factor and $T_S$ is the total system noise temperature.
Therefore $B$, $V$, and $T_S$ are crucial detector design parameters that determine the performance of the axion haloscope.
\par The CAPP18T haloscope in this report takes advantage of the $B^4$ dependence of the scan speed by using an 18\,T magnet, the strongest magnet among the existing haloscopes. The magnet is designed with a newly developed high-temperature superconductor (HTS) technology. A Josephson parametric converter (JPC) is used as the first stage rf signal amplifier to maintain system noise temperature below subkelvin order. Other significant components are a dilution refrigerator (DR), a field cancellation coil (FCC), and a frequency tunable copper cavity. This Letter reports the first results of the invisible axion dark matter search using the CAPP18T haloscope.

\begin{figure}[t!]
\centering
\includegraphics[width=0.98\linewidth]{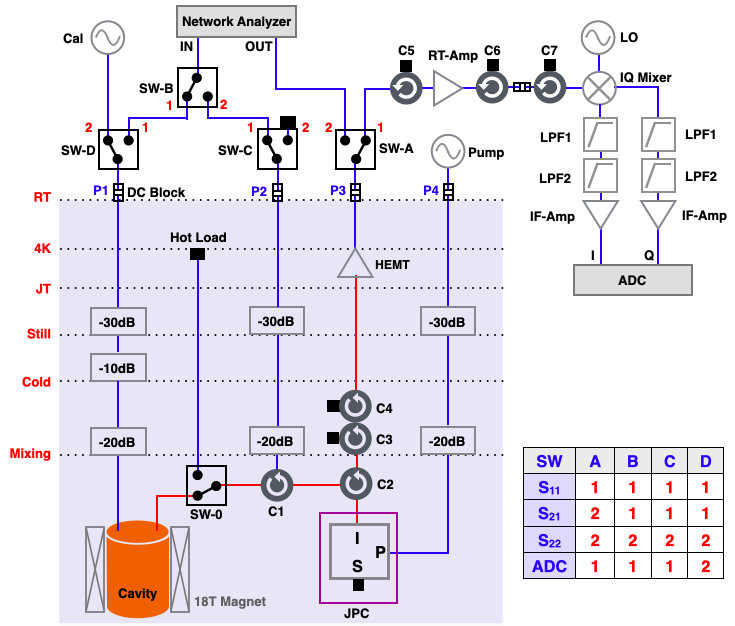}
\caption{ Schematic diagram of the CAPP18T rf receiver chain. The inset table shows the combinations of the room-temperature (RT) switches used to measure rf responses. The circulators ({\it C}1--{\it C}7), the amplifiers (HEMT, RT-amp, and IF-amp), the low pass filters (LPF1 and LPF2), and the switches (SW-0, {\it A}, {\it B}, {\it C}, and {\it D}) are represented by the corresponding symbols, respectively. The cryogenic rf switch (SW-0) can be toggled between the cavity and the hot load for noise calibrations and $Y$-factor measurements.}
\label{fig:rf}
\end{figure}
\par A schematic diagram of the CAPP18T rf receiver chain is shown in Fig.~\ref{fig:rf}. The 18\,T solenoid magnet consists of a stack of 44 double-pancake coils made of $\rm{GdBa_2Cu_3O_{7-{\it x}}}$ HTS tapes~\cite{hahn2011hts, hahn2013no, hahn2015construction, yoon201626t, jang2017design, kim2016400}. A typical operating current of the magnet is 199.2\,A at a central magnetic field of 18.0\,T. The total voltage of the magnet is below 9\,mV at normal operations. The stability of the magnetic field is better than 0.05\%. The axial magnetic fields [$B_z(r,z)$] at different radial and axial positions are measured and are consistent with the field simulation results in 1.0\%. Engineering details of the 18\,T HTS magnet can be found in Ref.~\cite{SuNAM2020a}.

\par A JPC, made by Quantum Circuits, Inc., is used as the first stage rf amplifier\,\cite{bergeal2010phase, Bergeal2010JPC, LiuJPC2017}. The JPC is a nondegenerate, phase-preserving parametric amplifier with quantum-limited added noise, which consists of a loop of four identical Josephson junctions. The JPC has three ports: signal ({\it S}), idler ({\it I}), and pump ({\it P}). An injected signal of the frequency near one of the resonant frequencies (signal) is amplified at that frequency. At the same time, the signal is amplified and converted into the frequency near the other resonant mode (idler). The pump tone at the sum frequency of the signal and idler provides energy for the amplification. The resonant frequency of the JPC is tuned using the flux coil under the JPC casing. The dynamic frequency ranges of the JPC are from 4.757 to 5.01\,GHz on the idler and from 7.720 to 8.802\,GHz on the signal. The pump tone frequency of the JPC is about 13.2\,GHz. In this experiment, the Idler mode is used for signal amplification. The typical gain of the JPC during the experiment is about 27\,dB. The JPC is installed at the mixing chamber stage. The added noise temperature of the JPC is measured to be $\sim$502\,mK at the base temperature of 60\,mK, which is consistent with the vendor specification and other measurements\,\cite{Roch2012,Bergeal2012}. This noise temperature is higher than that of an ideal parametric amplifier, which may be due to (1) the attenuation between the Josephson junctions and HEMT, (2) the imperfectly matched complex impedance of the junction\,\cite{Roch2012,Spietz2010}, and (3) the correlation between the gain and noise temperature due to the imperfect isolation of the circulators\,\cite{Bergeal2012}. The stray magnetic field from the 18\,T magnet at the location of the JPC was about 630\,G. This strong magnetic field is suppressed using the FCC, which otherwise can malfunction the JPC. A Hall sensor is installed about 1\,cm above the JPC and is operated in low current mode (4\,mA) to minimize Joule heating. During the axion search experiment, the magnetic field at the Hall sensor was maintained below 10\,G. No significant change in the frequency, noise temperature, or gain of the JPC associated with the residual magnetic field is observed.
\begin{figure}[t!]
\centering
\includegraphics[width=0.55\linewidth]{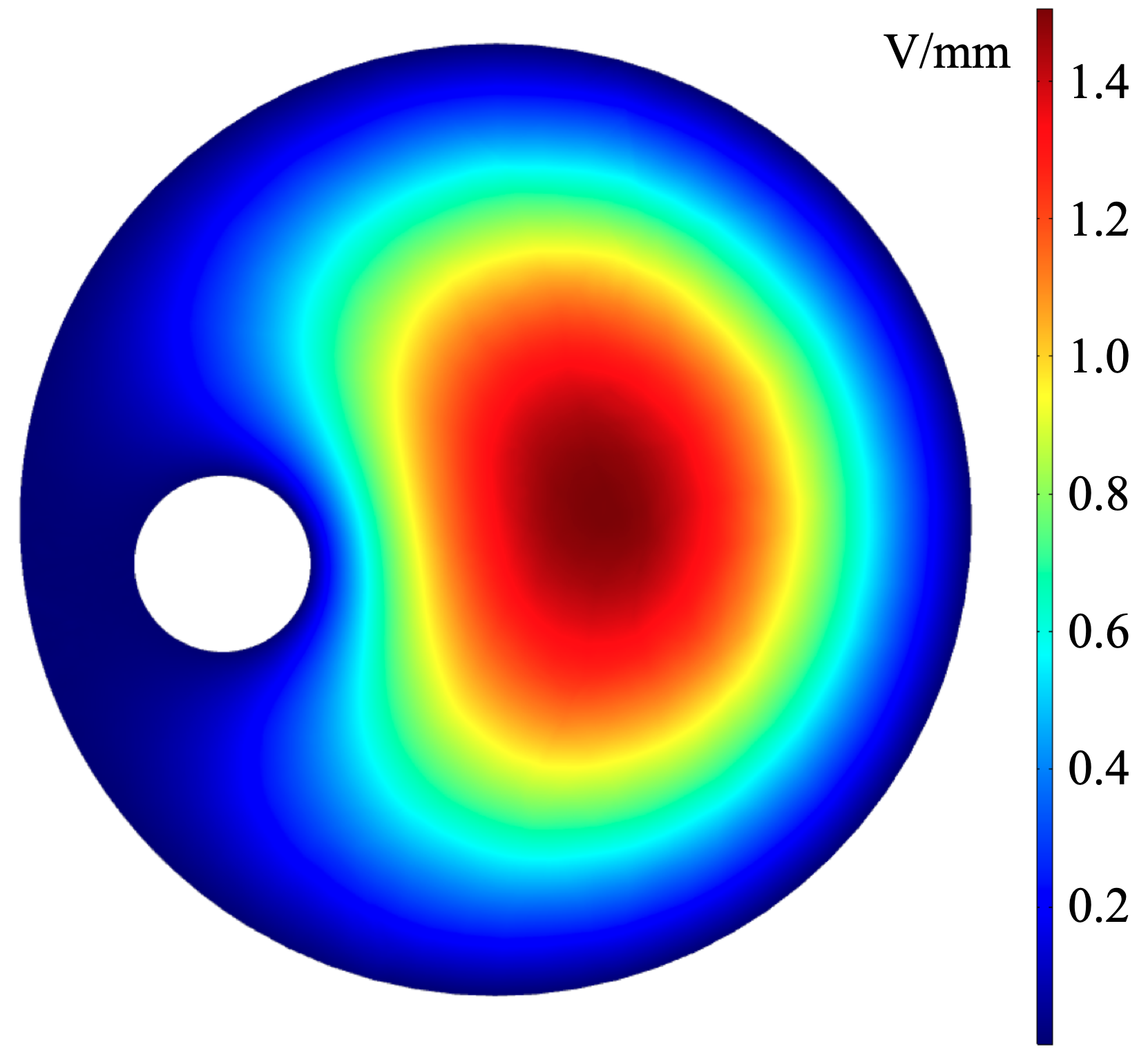}
\caption{A section view of an electric field map of a TM$_{010}$ mode at the center of the cavity. The resonant frequency is set by the position of the tuning rod.}   
\label{fig:efieldmap}        
\end{figure}
\par A cylindrical cavity with its inner diameter of 54.0\,mm and a height of 466.6\,mm is made of oxygen-free-high-conductivity (OFHC) copper with 99.99\% purity. The cavity is vertically split into two pieces. The resonant frequency ($\nu_C$) of the cavity is tuned by an off-centered cylindrical OFHC copper rod with a diameter of 10.0\,mm and a height of 465.6\,mm. Two stepper motors at room temperature are coupled to pulley systems in the DR using Kevlar wires. Each motor generates opposite directional motions to control the relative position of the tuning rod in the cavity. Another stepper motor is coupled to a linear motion pulley system to adjust the position of an rf coupling antenna. The average form factor of each frequency is calculated using a simulation software. The form factor varies from 0.56 to 0.58 in the frequency range of interest. Figure~\ref{fig:efieldmap} shows a section view of an electric field map at the center of the cavity. In the present tuning-rod configuration, four mode crossings between the TM$_{010}$ and TE or TEM modes are found. The crossing modes are at 4.840, 4.879, 4.948, and 5.042\,GHz, irrelevant to the current data-taking operation.

\par The rf signal in the cavity is transmitted to the strong coupling antenna and is propagated through the circulators ({\it C}1 and {\it C}2) to the {\it I} port of the JPC. Two circulators ({\it C}3 and {\it C}4) isolate the JPC and prevent backactions from the high electron mobility transistor (HEMT, model LNF-LNC2$\_$6A), the second-stage amplifier. The gain of the HEMT is about 37\,dB with an added noise temperature of 2\,K in the frequency region of interest. The rf signal is transferred to the transistor amplifiers at room temperature (RT amps). The noise contribution by the RT-rf chain to $T_S$ is negligible. A vector network analyzer (VNA) is used to characterize the receiver chain and the rf circuits. For rf chain calibrations, swept tones produced by the VNA are directed via RT switches to measure $\nu_C$, $Q_L$, $\beta$, and the gain of the JPC. To minimize signal loss, superconducting coaxial cables are used in the receiver chain from the cavity to the JPC, and the JPC to the HEMT. Frequency-dependent transmission losses of the rf chains are measured in a dedicated cryogenic setup. The measured transmission losses are used in the data analysis. The net attenuation between the strong antenna to the JPC is $-0.9$\,dB and is reflected in the magnitude of the signal power. The attenuation from the JPC output port to the HEMT input port is $-1.0$\,dB. For rf calibration, a $-84$\,dBm signal at $\nu_C$ is injected for a few seconds, and then a $\nu_C$+250\,kHz signal is injected during the data taking operation. An {\it I}/{\it Q} mixer down-converts the rf signals to intermediate frequency (IF) signals by superimposing the rf signal with a local oscillator (LO). The LO frequency is set 2.25\,MHz higher than the resonant frequency of the cavity to avoid unwanted ambient noise peaks in the IF band and to prevent interference with the analysis frequency band ($\Delta\nu_c$). The I (in-phase) and Q (quadrature-phase, 90$^\circ$ phase shift) signals keep the phase information of the input rf signal. These {\it I} and {\it Q} signals are transferred to a two-channel analog-to-digital-converter (ADC).

\par The front-end ADC board is a Signatec PX14400A (14-bit and 400\,MS/s of maximum sampling rate). The ADC has a 512\,MB on-board RAM, which allows first-in-first-out continuous data acquisition. All signal generators, VNA, and ADC board are synchronized with a standard 10\,MHz reference clock. The input {\it I} and {\it Q} signals are converted to power spectra using a fast-Fourier-transform (FFT) algorithm. An image rejection process is performed using the {\it I}/{\it Q} power spectrum. The rejection power of the imaginary band is better than 20\,dB at all IF bands. Multithread processors are used to compute FFTs in real-time data stream. A dead-time free condition of the DAQ is achieved in a sampling rate below 50\,MS/s; note that the data-taking rate of the CAPP18T axion search is 40\,MS/s ($2\times$20\,MS/s for {\it I} and {\it Q} channels). The sampling length of the time domain trace is 100\,ms, equivalent to 10\,Hz spectral resolution, which is good enough to probe the 5\,kHz of the expected bandwidth of an axion signal. 
Every 50 sets of the FFT-ed power spectra with a total of 5\,s long data are averaged out and written on the disk. In parallel, the first set of every 50 time-domain samples is stored for data quality tests.

\par Dark matter axion search data were collected from 30 November to 24 December 2020 in the frequency range from 4.7789 to 4.8094\,GHz. The $\nu_C$ and $Q_L$ of the cavity are measured by injecting an external rf signal through a weakly coupled antenna on the cavity. The coupling of the antenna to the cavity mode is measured by $S_{22}$. The average of the measured $Q_L$ is about 24\,400, with a standard deviation of 12.3\,\%. The typical measured value of $\beta$ is 1.02 with a standard deviation of 8.8\,\%. An IF power spectrum centered on the cavity resonant frequency is constructed by integrating the raw spectrum for 5\,s. The measurement duration of the power spectrum at a given frequency setup is varied from 10 min to several hours. The net exposure time at a given frequency mainly depends on the stability of the resonant frequency. If the gain of the JPC is off by 0.5\,dB from the peak gain at $\nu_C$, the resonant frequency of the JPC is readjusted to match $\nu_C$. The JPC gain curve is measured in every retuning process, and is used to scale the power spectrum.

\begin{figure}[t!]
\centering
\includegraphics[width=1.0\linewidth]{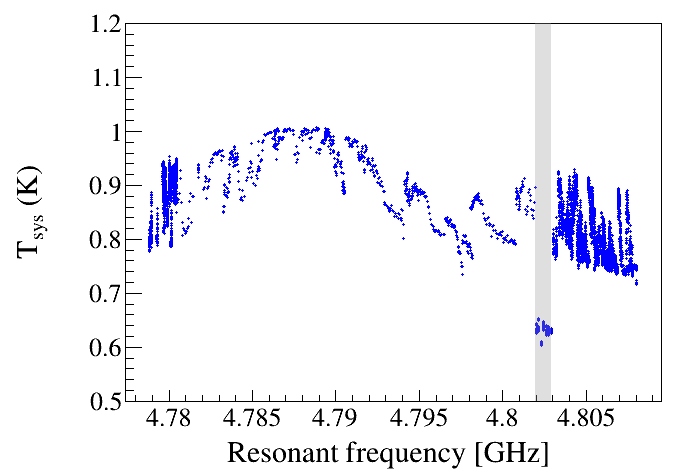}
\caption{Measured system noise temperatures at $\nu_C$. The gray band frequency region was measured in early August 2021 during the rescan operation. Note that the thermal condition of the detector is improved in the new setup.}
\label{fig:Tsys}
\end{figure}

\par The temperature of the DR mixing chamber was maintained at 50$-$60\,mK, and the temperature of the cavity at 100$-$120\,mK. A $Y$-factor measurement and a signal-to-noise ratio improvement (SNRI) method are adopted to calibrate the noise temperature of the system, as outlined in Ref.\,\cite{Du:2018uak}. $Y$-factor measurements are periodically carried out to monitor the system noise temperature. The JPC pump tone is unpowered during the measurements. The cold-noise power is measured by keeping the SW-0 to the cavity chain. For the hot-noise measurement, the SW-0 is toggled to the hot-load chain. The hot-load is a 50\,$\Omega$ rf terminator, anchored to the 4\,K plate. The physical temperature of the hot load varies from 4.4 to 5.3\,K, which depends on the operation cycle of the DR. The measured Johnson noise quanta at the HEMT input port without the JPC amplification is 19.31$\pm$0.66, which is stable within the error during the operation. The system noise temperature is given by $T_S=T_\text{HEMT} [(G_\text{off} P_\text{on})/(G_\text{on} P_\text{off})] ={T_\text{HEMT}}/{\text{SNRI}}$, where $T_\text{HEMT}$ is the noise temperature of the HEMT and downstream components, $G_\text{on/off}$ is the transfer function, and $P_\text{on/off}$ is the power at the target frequency for the JPC power on or off, respectively. The attenuation between the JPC and the HEMT is corrected in the gain estimation. The SNRI is defined as $(G_\text{on} P_\text{off})/(G_\text{off} P_\text{on})$.
Figure\,\ref{fig:Tsys} shows the measured $T_S$ as a function of $\nu_C$. We found that the SNRI measurements in the frequency region of 4.801\,960 to 4.802\,945\,GHz were unstable in the December 2020 data. This frequency region was scanned again during the August 2021 operation. 

\par The axion search was carried out in a total of 24.5 calendar days. Only the data in good detector conditions are included in the data processing. The instability of the detector system caused by the detector vibration is the primary source of inefficiency. Especially the LHe transfer to the DR causes the most significant and long-lasting ($\sim$4 h) detector vibrations. The preselection process of good data leaves 1\,116\,404.4\,s (12.9\,days) of net amount of data. Anomalous SNRIs are observed in the frequency range of 4.801\,960 to 4.802\,945\,GHz. These data samples are removed from the initial analysis, which leaves 1\,063\,957.5\,s (12.3\,days). To select only the data under stable $\nu_C$ conditions, data belonging to $|\nu_C(t_{i+1})-\nu_C(t_{i})| \ge 10$\,kHz are removed, where $\nu_C(t_{i})$ is the resonant frequency measured at $t_i$. This leaves 1\,060\,688.8\,s (12.3\,days) of data. The stability of $Q_L$ is also required by removing data with $|Q_L(t_i)-Q_L(t_{i+1})|\ge$\,1\,000, which leaves 1\,047\,782.9\,s (12.1\,days) of data. These data samples are further processed to search for the axion dark matter signals.

\begin{figure}[t!]
\centering
\includegraphics[width=1.0\linewidth]{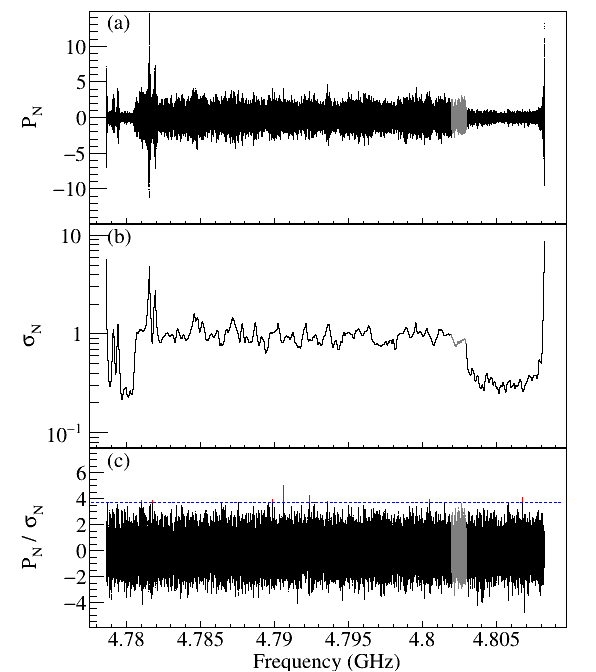}
\caption{Normalized grand power spectrum and uncertainty. (a) Normalized power spectrum ($P_N$). (b) Uncertainty of the normalized power spectrum at each frequency ($\sigma_N$). (c) Scaled grand power spectrum ($P_N/\sigma_N$). The blue dotted line shows the 3.718\,$\sigma$ threshold. Signals that exceed the threshold are indicated in red color above the blue dotted line. The gray data points were measured in early August 2021. }
\label{fig:pn}
\end{figure}

\par The axion search data analysis is carried out using the method outlined in Ref.\,\cite{Brubaker:2017rna} to properly compare the results from other experiments. The Savitzky-Golay (SG) filter\,\cite{SGfilter1964} is applied to remove the characteristic spectral structure in each raw power spectrum while keeping the clustered spectral excess on the scale of $<\Delta \nu_a$. A polynomial function of degrees 3 and 5\,001 data points are used as the SG-filter parameters. The raw power spectrum is normalized using the SG-filter function to obtain a dimensionless spectrum. The mean of the normalized spectrum is shifted to zero. The attenuation of the axion signal to noise power by the SG-filter is estimated to be $\varepsilon_{_\text{SG}}=0.888$. The spectrum is scaled using the measured $T_S$, $Q_L$, rf gains, attenuation factors, and Lorentzian axion conversion power profile. 

\par The expected power of the hypothetical dark matter axion is expressed as $P = \xi(\nu) P^a / \cal{L}(\nu)$, where $\xi(\nu)$ is the measured signal attenuation (about $-$0.9\,dB) from the strong antenna to the JPC, and $\cal{L}(\nu)$ is the normalized Lorentzian curve; ${\cal L}(\nu)=1+4{(\nu - \nu_{c})^2}/{(\Delta \nu_{c})^2}$. Each bin in the rescaled spectrum forms an independent normal distribution with different standard deviations centered at zero. Total of 5906 sets of spectra are combined to form a grand power spectrum. Each frequency bin is weighted by the inverse of the corresponding variance. A single grand combined spectrum $P_N$ is constructed over the whole scanned frequency range by taking weighted mean of all rescaled spectra. Figure\,\ref{fig:pn} shows the grand power spectrum, the weighted standard deviation ($\sigma_N$) of $P_N$, and the scaled grand power spectrum. The axion signal search is performed on the scaled grand power spectrum ($P_N/\sigma_N$). A running search window method is used to reduce binning bias effects; a search window of 5\,kHz band is set, and the center frequency of the window is shifted 10\,Hz for each test over the whole frequency range. The central frequency of the 5\,kHz window is taken as the representative frequency. Peaks exceeding the $3.718\,\sigma$ threshold are tagged as the potential candidates of the dark matter axion signal at 90\% confidence for 5.0\,$\sigma$ excess singal-to-noise level target\,\cite{Brubaker:2017rna}. A total of eight candidates exceeding the threshold are found [see Fig.\,\ref{fig:pn} (c)]. Likelihood tests are carried out for all excess data points with the boosted Maxwell-Boltzmann (MB) model~\cite{PhysRevD.42.3572}. Among them three candidates are consistent with the random fluctuations ($>$\,64\%). 
\begin{figure*}[t!]
    \includegraphics[width=0.96\linewidth]{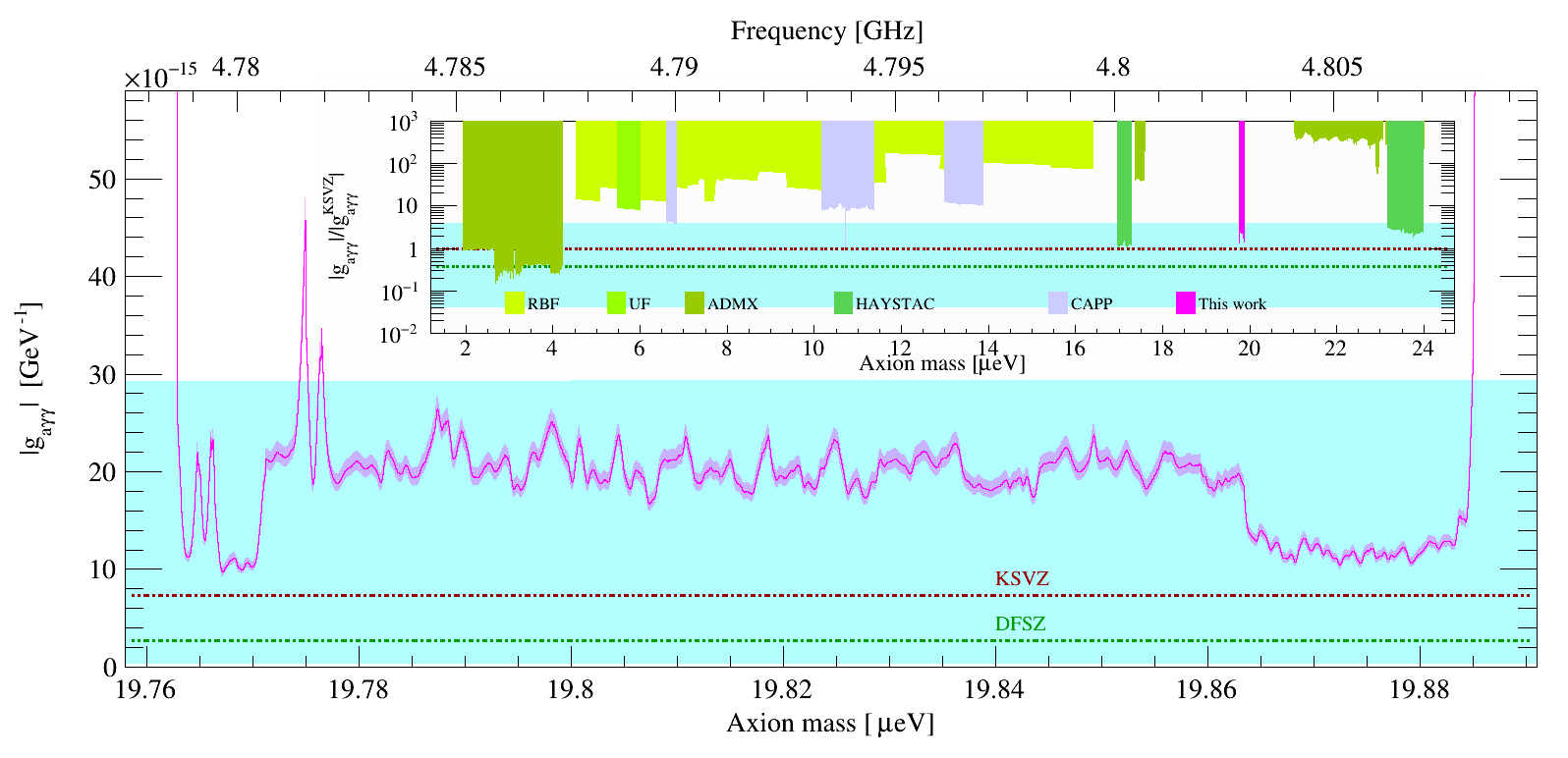}
    \caption{Exclusion limit of the first CAPP18T axion dark matter search. The magenta curve shows the exclusion limit of the MB model at 90\% CL from this work. The light magenta-band at the exclusion curve boundary represents the systematic uncertainty. The KSVZ (DFSZ) model is shown as a dotted-brown (dotted-green) line. The light cyan band shows the uncertainty of the invisible axion models\,\cite{Cheng1995a}. The inset shows the limit from this work (magenta) together with previous haloscope limits from ADMX (olive, Refs.\,\cite{Hagmann1998a,Asztalos:2001tf,Asztalos2002a, Asztalos:2003px, Asztalos:2009yp,Sloan2016a, Boutan2018a,ADMX:2021nhd}, CL\,=\,95\%), HAYSTAC (green, Refs.\,\cite{Brubaker:2016ktl, Backes2021a, PhysRevD.97.092001}, CL\,=\,90\%), and CAPP (light purple, Refs.\,\cite{Lee:2020cfj, Jeong:2020cwz, Kwon2021a}, CL\,=\,90\%). RBF (lime, Refs.\,\cite{DePanfilis1987a, Wuensch1989a}, CL\,=\,95\%) and UF (light green, Ref.\,\cite{Hagmann1990a}, CL\,=\,95\%) limits shown here are rescaled based on $\rho_a$\,=\,0.45 GeV/cm$^3$.}
\label{fig:limit}
\end{figure*}
These three signals show localized sharp peaks in a narrow frequency range. The rest of the five candidates are consistent with the MB model better than 1\% of probability. These five data points are classified as the {\it rescan candidates}. The center frequencies of the candidates are 4.789\,855, 4.790\,596, 4.792\,344, 4.793\,603, and 4.806\,745\,GHz. 

\par The rescan experiment was carried out in early August 2021 after the relocation of the CAPP18T set up at a new experiment site. The five candidate frequencies were scanned with more than a factor of 2 exposure time than the original data. The rescan data rule out all five candidates. These rescan data are not included in the axion search limit calculation. 

\par The absence of any significant axion signal in the data rules out the axion dark matter model parameters as shown in Fig.\,\ref{fig:limit}. In the mass range of 19.772 to 19.863\,$\mu$eV shallow-scan region, the results set the upper bound of 2.7 $\times|g_{a\gamma\gamma}^{\text{KSVZ}}|$ at 90\% confidence level (CL). In the mass range of 19.764 to 19.771\,$\mu$eV (19.863 to 19.890\,$\mu$eV) deep-scan region, the results set the upper bound of 1.5$\times|g_{a\gamma\gamma}^{\text{KSVZ}}|$ (1.7$\times|g_{a\gamma\gamma}^{\text{KSVZ}}|$) at 90\% CL. The four insensitive ``notches" at 4.77912, 4.77947, 4.78159, and 4.78191\,GHz are caused by the system instabilities during the data taking. The data samples of these frequency regions are removed in the data selection process. Systematic uncertainties are evaluated as a function of frequency on the expected axion signal power from the cavity. Primary sources of the systematic uncertainties are $B^2 V$ (1.4\%), $Q_L$ (0.5\%), coupling ($\beta$, 0.4\%), form factor ($C_{010}$, 3.9\%), and $T_S$ (8.5\%). The total systematic uncertainty on the expected signal power is 9.5\%. 
\par We report the first results of the invisible axion dark matter search using the CAPP18T axion haloscope. No significant signal consistent with the Galactic halo dark matter axion is found. The results set the best upper bound of the axion-photon-photon coupling and exclude the invisible axion model parameters in the mass ranges of 19.764 to 19.890\,$\mu$eV at 90\% CL. This remarkable sensitivity is achieved by the combination of the strongest magnetic field among the existing haloscope experiments and achieving a low-noise amplification of microwave signals using a JPC.

\par This research is supported by the Institute for Basic Science (IBS-R017-D1-2021-a00/IBS-R017-G1-2021-a00). This work is also supported by the New Faculty Startup Fund from Seoul National University. We also thank the team of the Superconducting Radio Frequency Testing Facility at Rare Isotope Science Project at IBS for sharing the he helium plant and the experiment space. 
\bibliography{capp18t2021}
\end{document}